\documentclass[aps,prx,reprint,superscriptaddress,floatfix]{revtex4-2}
\usepackage{amsmath,amssymb,amsthm,bm,graphicx,booktabs,xcolor,multirow,array}
\usepackage{tikz}
\usetikzlibrary{quantikz,positioning,arrows.meta}
\usepackage[colorlinks=true,linkcolor=blue,citecolor=blue,urlcolor=blue]{hyperref}
\makeatletter\g@addto@macro\UrlBreaks{\do\_\do\-}\makeatother
\newtheorem{theorem}{Theorem}
\newcommand{\Ha}{\ensuremath{\,\mathrm{Ha}}}
\newcommand{\mHa}{\ensuremath{\,\mathrm{mHa}}}
\providecommand{\ket}[1]{|#1\rangle}
\providecommand{\bra}[1]{\langle #1|}
\providecommand{\proj}[1]{|#1\rangle\langle #1|}
\newcommand{\brot}{b_{\mathrm{rot}}}
\newcommand{\FeS}{Fe$_2$S$_2$}
\newcommand{\FeSS}{Fe$_4$S$_4$}
\newcommand{\CNOT}{\mathrm{CNOT}}
\begin{document}

\title{Exact verification of Givens rotation circuits for efficient quantum phase estimation of iron--sulfur clusters}

\author{Lizhong Fu}
\affiliation{State Key Laboratory of Precision and Intelligent Chemistry, Hefei National Laboratory, University of Science and Technology of China, Hefei, China}
\author{Honghui Shang}
\email[Corresponding author: ]{shanghui.ustc@gmail.com}
\affiliation{State Key Laboratory of Precision and Intelligent Chemistry, Hefei National Laboratory, University of Science and Technology of China, Hefei, China}

\date{October 28, 2026}

\begin{abstract}
Givens rotations with angles stored in quantum registers account for much of the cost of published
quantum phase estimation algorithms based on double-factorized tensor hypercontraction and spectral amplification.
These rotations create superpositions and cannot be verified directly by the basis-state evaluators used to
optimize classical reversible circuits. We give an exact verification method that separates out a fixed Clifford
change of basis and tracks basis states and phases in the remaining circuit. Integer conditions establish correctness of the full
coherent action, including measurement-based uncomputation and restoration of the phase-gradient resource state.
Under our interpretation of Low \emph{et al.}'s angle grid and rotation convention, simplifying the adders lowers
the rotation cost relative to the published accounting; fusing the two additions into one masked addition gives
an additional saving. We verify both implementations exhaustively at every precision used in the published walks
and prove the fused circuit correct for all widths. For benchmarks including FeMoco and other iron--sulfur clusters,
replacing only the rotation term lowers total phase-estimation Toffoli counts by 9.5--14.3\% under the published
error budget, with the walk operator unchanged.
\end{abstract}

\maketitle

\section{Introduction}\label{sec:intro}

Computing the ground-state energy of an active-space model of a metalloenzyme cofactor, such as the iron--molybdenum
cofactor (FeMoco) of nitrogenase or an iron--sulfur cluster, is the flagship application of fault-tolerant quantum
chemistry~\cite{Reiher2017,Li2019,vonBurg2021,Lee2021,LeeEvidence2023,Low2025}. The standard algorithm is quantum phase
estimation (QPE)~\cite{Kitaev1997} on a qubitized walk operator~\cite{Babbush2018,LowChuang2019}. Its Toffoli cost is the number of walk steps multiplied by the number of Toffolis per step.
The step count grows with the normalization $\lambda$ of the Hamiltonian's block encoding. The lowest published estimates that we know of, by Low \emph{et al.}~\cite{Low2025},
combine a
double-factorized tensor-hypercontraction (DFTHC) representation with block-invariant symmetry shifts
(BLISS)~\cite{Loaiza2023} and spectral amplification (SA). Under the error budget used in their Table~V, they reach
$3.99\times10^{7}$ Toffolis for the \FeS{} cluster of Li and Chan~\cite{LiChan2017} and $1.00\times10^{9}$ Toffolis
for the 76-orbital FeMoco model of Li \emph{et al.}~\cite{Li2019}. For the latter, they state 8.6 hours on 4.5 million
physical qubits alongside the preprint's $9.99\times10^{8}$ Toffolis (Appendix~\ref{app:cost}).

In this walk, 44--72\% of the Toffolis of each published step implement networks of Givens rotations whose angles are
loaded from a lookup table into quantum registers. Two questions about these rotations bear directly on the published
costs. The first is their price. Low \emph{et al.}'s totals, their cost breakdown and the costing code released with
their paper~\cite{Low2025code} charge $2\brot$ Toffolis per rotation, where $\brot$ is the angle precision, but an
equation, a table and a passage of their appendices charge $\brot$. The lower count is the price of a single
phase-gradient addition, which applies a single Pauli rotation. That rotates only one of the two Majorana operators of
each mode, whereas their select step needs both, and we show that one signed addition of the angle register, if nothing
else reads the angle register or alters the phase-gradient register, cannot implement a Givens rotation
(Sec.~\ref{sec:cost}, Appendix~\ref{app:oneadd-proof}). Our baseline is therefore the $2\brot$ of their totals and code, and every improvement claimed
below is relative to it; if a complete Givens rotation could be done for $\brot$ Toffolis, our constructions would not
improve on it, but we know of no such construction.

The second question is verification. A coherent circuit that is wrong on a small fraction of inputs can spoil a
computation that applies it $10^{4}$ to a few $10^{5}$ times, so a cheaper rotation is useful only if it is verified exactly.
Basis-state evaluators follow one computational basis state at a time, exhaustively when the input space is small
and by sampling otherwise. Gu \emph{et al.}~\cite{Gu2026} used such an evaluator in an evolutionary search to score
circuits proposed by a large language model (LLM) and written in the kickmix circuit model~\cite{Babbush2026kick}.
Such evaluation is exact for circuits that map basis states to basis states up to
a sign, which a Givens rotation does not. Gu \emph{et al.} note this limitation and, for their chemistry task, the same \FeS{} cluster,
optimized the sparse Pauli representation~\cite{Babbush2018,Berry2019} instead, reducing the data-dependent part of its
walk step from about 55\,000 to 15\,000 Toffolis.

Our first result extends exhaustive verification to these rotations. Theorem~\ref{thm:verify} gives four integer
conditions for circuits with a fixed change of basis, a basis-permuting core with $\mathbb{Z}_4$ phases, and a
phase-gradient resource state. The conditions are necessary and sufficient for implementing the target diagonal action
in the transformed basis when every rescaled measurement branch must match exactly, including its global phase. The certificate
combines standard ingredients: $\mathbb{Z}_4$ phases extend sign tracking, the phase-gradient state is an
eigenvector of modular addition, and measurement-based uncomputation erases temporaries. After factoring out a fixed change of
basis, these ingredients give a complete certificate for a class that contains rotations by loaded angles,
their controlled versions and Givens rotations. The class includes exact circuits that a comparison with the
textbook adder's output rejects: the two cheapest rotation circuits below leave the gradient register in a
different state on 3/8 of all basis inputs and are correct only because a phase compensates
(Sec.~\ref{sec:adds}).

\begin{figure*}[t]
\centering
\makebox[\textwidth][l]{\small\textbf{(a)}}\\[2pt]
\begin{tikzpicture}[box/.style={draw=black!70,rounded corners=2pt,align=center,font=\footnotesize,minimum height=1.25cm,
  text width=3.05cm,inner sep=3pt},arr/.style={-{Stealth[length=2.2mm]},thick,black!70},lab/.style={font=\scriptsize,
  align=center,text width=1.35cm}]
\node[box,fill=blue!6] (a) {Givens rotation $G(\theta)$:\\ two commuting Pauli rotations,\\ priced by Low \emph{et al.}
at $2\brot$ Toffolis};
\node[box,fill=blue!6,right=1.5cm of a] (b) {one rotation $e^{i\theta Z_p}$\\ controlled by the parity qubit $q$\\
(Eq.~(\ref{eq:frame}))};
\node[box,fill=red!6,right=1.5cm of b] (c) {one addition of the masked angle $q\wedge v$ into $\ket{G_b}$, $b=\brot-1$:\\
$2\brot-5$ Toffolis};
\node[box,fill=green!6,right=1.5cm of c] (d) {core permutes basis states, phases $i^{k}$:\\ an exhaustive check verifies
it (Theorem~\ref{thm:verify})};
\draw[arr] (a) -- node[lab,above]{basis change $F$} (b);
\draw[arr] (b) -- node[lab,above]{grid $2^{\brot-1}$, kickback} (c);
\draw[arr] (c) -- node[lab,above]{core between $F$ and $F^\dagger$} (d);
\end{tikzpicture}

\vspace{0.2cm}
\makebox[\textwidth][l]{\small\textbf{(b)}}\\[-4pt]
\begin{quantikz}[row sep={0.6cm,between origins}, column sep=0.22cm, font=\small]
\lstick{$p$} & \ctrl{1}\gategroup[2,steps=3,style={dashed,rounded corners,fill=blue!6,inner xsep=1pt},background,label style={label position=above,anchor=south,yshift=-0.1cm}]{$F$} & \gate{S^\dagger} & \gate{H} & \qw\gategroup[7,steps=4,style={dashed,rounded corners,fill=red!5,inner xsep=1pt},background,label style={label position=above,anchor=south,yshift=-0.1cm}]{\shortstack{core: $2b-3$ Toffolis\\peak temporaries: $2b-3$}} & \ctrl{4} & \qw & \qw & \gate{H}\gategroup[2,steps=3,style={dashed,rounded corners,fill=blue!6,inner xsep=1pt},background,label style={label position=above,anchor=south,yshift=-0.1cm}]{$F^\dagger$} & \gate{S} & \ctrl{1} & \qw \\
\lstick{$q$} & \targ{} & \qw & \qw & \ctrl{1} & \qw & \ctrl{2} & \ctrl{1} & \qw & \qw & \targ{} & \qw \\
\lstick{$v_{0..b-2}$} & \qwbundle{b-1} & \qw & \qw & \ctrl{2} & \qw & \qw & \ctrl{2} & \qw & \qw & \qw & \qw \\
\lstick{$v_{b-1}$} & \qw & \qw & \qw & \qw & \qw & \control{} & \qw & \qw & \qw & \qw & \qw \\
\lstick{$\ket{0}^{\otimes b-1}$ (masks)} & \qwbundle{b-1} & \qw & \qw & \gate{\wedge} & \gate[2]{\substack{\pm m\ \mathrm{mod}\ 2^{b}\\ b-2\ \text{carry qubits}}} & \qw & \gate{\mathcal{M}_X} & \qw & \qw & \qw & \rstick{$\ket{0}^{\otimes b-1}$} \\
\lstick{$\ket{G_b}$} & \qwbundle{b} & \qw & \qw & \qw & & \qw & \qw & \qw & \qw & \qw & \rstick{$\ket{G_b}$} \\
\lstick{$v_{b}$ (worth $2\pi$)} & \qw & \qw & \qw & \qw & \qw & \qw & \qw & \qw & \qw & \qw & \qw
\end{quantikz}
\caption{\label{fig:gadget}The fused Givens rotation. (a)~The three steps from a Givens rotation priced as two additions
to an exhaustively checkable circuit. (b)~The circuit for $G(2\pi v/2^{b})$ on adjacent Jordan--Wigner modes $p,q$. The
Clifford unitary $F=H_pS_p^\dagger\CNOT_{p\to q}$ turns $G(\theta)$ into $\proj{0}_q\otimes1+\proj{1}_q\otimes e^{i\theta
Z_p}$. In the transformed basis, $b-1$ temporary logical ANDs ($\wedge$) form the masks $m_i=q\wedge v_i$; one ripple-carry
addition with $b-2$ temporary-AND carries adds $\pm m$ into the phase-gradient register $\ket{G_b}$ (minus when $z_p=1$,
through CNOT layers from $p$), with the two top gradient bits replaced by $S^{\pm1}$ gates on the bit and carry that would
enter them; the top mask bit enters only as the phase $\mathrm{CZ}(q,v_{b-1})$. Masks and carries are erased by $X$-basis
measurements with CZ fix-ups ($\mathcal{M}_X$). The angle bit $v_b$, worth $2\pi$ on Low \emph{et al.}'s grid, is never
touched. With $b=\brot-1$ the circuit uses $2\brot-5$ Toffolis, $2\brot-5$ measurements and $2\brot-5$ temporary qubits
at peak; the inverse rotation flips the sign. Appendix~\ref{app:listing} lists every gate.}
\end{figure*}
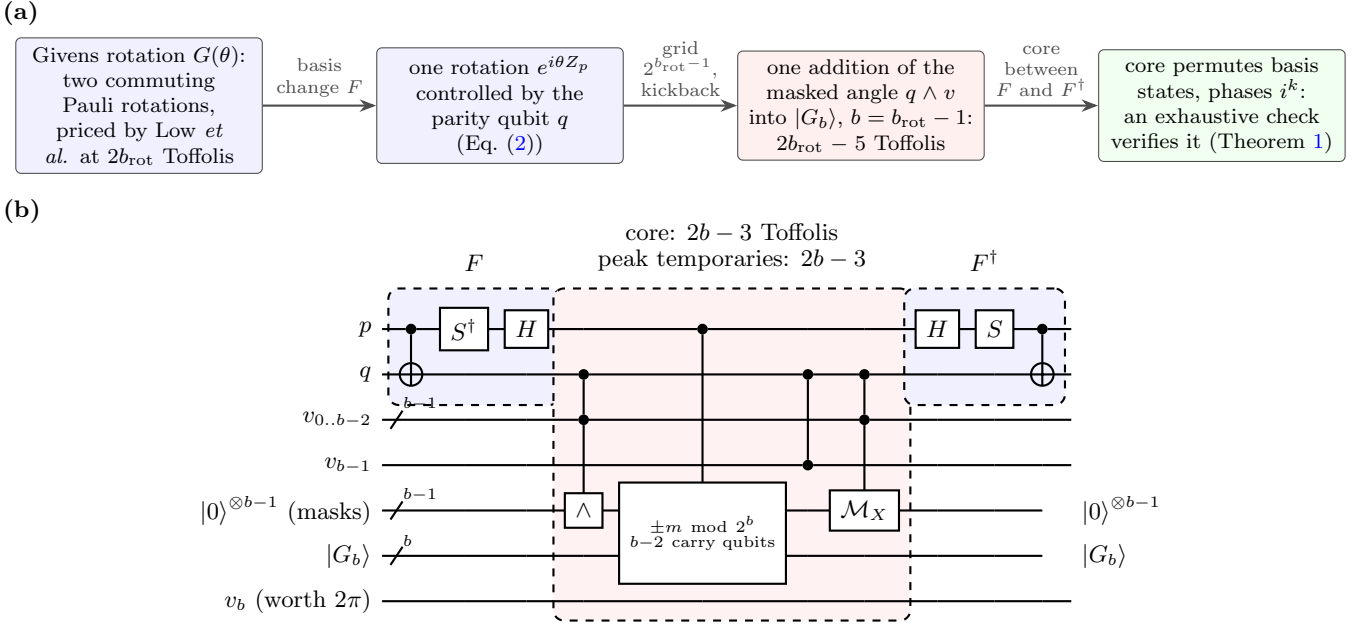

Our second result prices the rotations exactly (Fig.~\ref{fig:gadget}). A Clifford change of basis diagonalizes the two
commuting Pauli rotations of a Givens rotation~\cite{Arrazola2022}. With a quantum angle register, the two
phase-gradient additions then become one addition masked by a parity qubit. On our reading, Low \emph{et al.}'s
randomized rounding already places every angle on a $2^{\brot-1}$ grid, and the rotation is $2\pi$-periodic in
that angle, so the top angle bit never has to be read. With the top of the arithmetic replaced by Clifford
phases~\cite{Gidney2018,Sanders2020}, the rotation costs $2\brot-5$ Toffolis. Using two simplified adders
on the same grid already costs $2\brot-4$, so four of the five Toffolis saved per rotation are an accounting
correction and one is the fusion (Table~\ref{tab:ladder}). Every construction on this ladder is verified
exhaustively at $\brot=15$, the precision of the \FeS{} and FeMoco-76 walks; the rotations with simplified adders and the fused circuit
are verified at every width the published walks use, up to $2^{38}$ basis states, and the fused one is proved
for every width.

The formulas in Low \emph{et al.}'s released code reproduce the published costs of all 14 walk steps exactly.
We change only the rotation term, giving a matched comparison with the walk operator, $\lambda_{\mathrm{eff}}$,
the number of steps and the error budget unchanged. Toffoli counts fall by 9.5--14.3\% for the eight systems under their Table~V error budget and by 6.1--12.9\%
for the six of their statistical one, and the fusion accounts for 1.9--2.9 and 1.2--2.6 percentage points of these. For
\FeS{}, phase estimation with $\sigma_{\mathrm{PEA}}=1\mHa$ then needs $3.60\times10^{7}$ Toffolis ($3.06\times10^{7}$
with the statistical budget's $\sigma_{\mathrm{PEA}}=1.4\mHa$); for FeMoco-76, our transcription of the published surface-code model gives 7.81\,h instead
of 8.71\,h on the same hardware. Two readings of Low \emph{et al.}'s appendix, the angle grid and the angle convention,
are assumptions; if either fails, the fused construction one bit wider applies and saves 3 rather than 5 Toffolis per
rotation, 2 from simplified adders and 1 from the fusion.

The fused circuit was developed in interactive sessions with an LLM coding agent (see Acknowledgments), without
an automated search. Its correctness at every width is derived in Appendix~\ref{app:givens-proof}, and the evaluator independently
verifies the circuit at each width in use. Whether the evaluator can drive a search, as the kickmix evaluator
did for Gu \emph{et al.}, remains open (Sec.~\ref{sec:discussion}).

For \FeS{}, we also compare our estimates with the one implied by Gu \emph{et al.}'s optimized sparse walk for
the same integrals (Sec.~\ref{sec:sparse}). The two estimates use different representations with different
approximation errors. Their difference of almost two orders of magnitude comes from the representation and the
associated walk (DFTHC with BLISS and spectral amplification), rather than from either circuit optimization.

\section{Givens rotations with quantum angles}\label{sec:givens}

\subsection{The rotation and its published cost}\label{sec:cost}

The DFTHC+BLISS+SA walk of Low \emph{et al.} applies, in each of its two SEL calls, a basis rotation $U_{\mathrm{Rot}}$ and
its inverse, each a chain of $N-1$ Givens rotations on adjacent Jordan--Wigner modes of $N$ spatial
orbitals~\cite[App.~B\,5]{Low2025}, with the $N-1$ angles loaded from a lookup table into quantum registers. A walk step
therefore contains $4(N-1)$ rotations. On adjacent modes $p,q$,
\begin{equation}
G(\theta)=\exp\!\bigl[-\tfrac{i\theta}{2}(X_pY_q-Y_pX_q)\bigr]
=e^{-\frac{i\theta}{2}X_pY_q}\,e^{+\frac{i\theta}{2}Y_pX_q},
\label{eq:givens}
\end{equation}
a product of two commuting Pauli rotations with the same angle, which rotates the Majorana vector of the pair by
$\theta$. Each factor can be applied as one addition of the angle register into a phase-gradient register
$\ket{G_b}=2^{-b/2}\sum_{y}e^{-2\pi iy/2^b}\ket{y}$, which multiplies the state by $e^{2\pi iT/2^b}$ when $T$ is added.
Low \emph{et al.} use such additions (their App.~B\,7), and their counts correspond to $\brot$ Toffolis per $\brot$-bit
addition. Their App.~B\,5 (``each $U_{\mathrm{Rot}}$ can be implemented with $2(N-1)\brot$ Toffoli gates''), their
Sec.~IV ($\approx4(N-1)\brot$ per application of SEL), their Table~III ($4\brot(N-1)$ per SEL), the ``Select Toffolis''
of their Tables~X and~XII ($8(N-1)\brot$ per step), the totals of their Tables~V and~XI, and the select cost of their
released code~\cite{Low2025code}, $4(N-1)\brot$ per SEL call, all charge $2\brot$ per Givens rotation, two additions.
This is our baseline.

Their Eq.~(B46) and Table~IX instead charge $2(N-1)\brot$ for the forward and reverse rotations of a SEL call together,
and their App.~D states that discretizing the angles that ``realize the Majorana rotation from $\gamma_{000}$ to
$\gamma_{\vec u00}$'' leads to ``a Toffoli cost of $(N-1)b_{\mathrm{bits}}$'': one addition per Givens rotation. The same
count follows from their description in App.~B\,5 of $U(\vec u)$ as ``comprised of $N-1$ different arbitrary single-qubit
rotations'' and $O(N)$ Clifford gates with $U^\dagger\gamma_{00x}U=\tilde\gamma_{\vec u0x}$. A single-qubit rotation
between Clifford gates is one Pauli rotation $e^{i\phi P}$, which is what one signed addition of the angle register
applies. But Low \emph{et al.} build $U_{\mathrm{Rot}}$ from Givens rotations (their App.~B\,5 and Eq.~(B46)), and their
Eq.~(B21) requires it to rotate both Majorana operators $\gamma_{00x}$, $x=0,1$, which their select step applies
(their App.~B\,6). A Givens rotation is the product of two Pauli rotations, one for each of the two sets of Majorana operators indexed by $x$
[Eq.~(\ref{eq:givens})]; a single one, such as $e^{i\phi X_pY_q}$, rotates the operators of one set and commutes with
those of the other, so a chain of them, one per angle, satisfies Eq.~(B21) for one value of $x$ only.
Appendix~\ref{app:oneadd-proof} gives the general argument for a single rotation: one signed addition of the angle register, if
no other gate reads the angle register or alters the gradient register, cannot implement $G(\theta)$, under any change of basis, with
any other ancillas and up to any global phase. If a complete Givens rotation could be done for $\brot$ Toffolis, the FeMoco-76 step would cost 10\,129 Toffolis instead of 14\,629, and the
constructions below would not improve on it; we know of no construction at that price.

\subsection{Diagonalization by a Clifford change of basis}

The Clifford unitary $F=H_pS^\dagger_p\CNOT_{p\to q}$ maps the generator of Eq.~(\ref{eq:givens}) to $-2Z_p\proj{1}_q$, so
\begin{equation}
F\,G(\theta)\,F^\dagger=\proj{0}_q\otimes 1+\proj{1}_q\otimes e^{i\theta Z_p}
\label{eq:frame}
\end{equation}
as derived in Appendix~\ref{app:givens-proof}. This is the familiar decomposition of a Givens rotation into CNOTs around a controlled
rotation~\cite{Arrazola2022}, here applied to the rotations of a basis-change network~\cite{Kivlichan2018,Jiang2018}. With
a quantum angle register the controlled rotation is a single phase-gradient addition of the angle masked by $q$~\cite{Caesura2025,Caesura2026Erratum}: the
control costs $b-1$ temporary ANDs, and the sign $(-1)^{z_p}$ costs two sets of CNOTs that turn the addition into a
subtraction. The mask ANDs read the angle register, so the single-addition restriction of Appendix~\ref{app:oneadd-proof} does not apply.

\subsection{The angle grid}

Low \emph{et al.}'s randomized rounding~\cite[App.~D]{Low2025} stores each angle as $v=x+B$, with
$x=\lfloor 2^{\brot-1}\theta/2\pi\rfloor$ and $B$ a random bit that is 1 with probability equal to the fractional part, so
every angle is a multiple of $2\pi/2^{\brot-1}$ and the rounding is unbiased. (Their prose describes $x$ as ``the first
$\brot-1$ bits'' and $B$ as ``the last bit''; only the stated formula $2\pi(x+B)/2^{\brot-1}$ gives unbiased rounding,
and we follow it.) A $\brot$-bit register is needed only because $v$ can equal $2^{\brot-1}$, which is the angle $2\pi$.
We read $\theta$ as the angle by which $U_{\mathrm{Rot}}$ rotates the Majorana vector, as their App.~D describes the
angles (``$\vec\theta\in[0,2\pi)^{N-1}$ that realize the Majorana rotation'') and as in Eq.~(\ref{eq:givens}); then $G$ is exactly $2\pi$-periodic (Appendix~\ref{app:givens-proof}), the circuit never reads the
top bit, and it runs on the $2^{b}$ grid with $b=\brot-1$. A lookup table that stores $x$ and $B$ separately can store
$(x+B)\bmod2^{b}$ instead, which changes classical data only. No further bit can be dropped, since
$G(\theta+\pi)=Z_pZ_q\,G(\theta)$.

\subsection{The fused circuit and the cost ladder}

The two most significant qubits of a phase-gradient register form an eigenstate of addition modulo 4, so instead of computing the last
carry the circuit applies $S^{\pm1}$ to the bit and the carry that would enter them, and the top mask bit becomes the
phase $\mathrm{CZ}(q,v_{b-1})$; this leaves $b-2$ carries, as in the usual $(b-2)$-Toffoli phase-gradient
rotation~\cite{Gidney2018,Sanders2020}. Every temporary is erased by an $X$-basis measurement and a CZ
fix-up~\cite{Gidney2018}, which is correct on both outcomes when the temporary still equals the AND of its inputs when it
is measured (Appendix~\ref{app:verification-proof}). The circuit of Fig.~\ref{fig:gadget}b uses $(b-1)+(b-2)=2b-3$ Toffolis, $2\brot-5$ on Low
\emph{et al.}'s grid.

In Table~\ref{tab:ladder}, two additions with full adders on a $2^{\brot}$ grid reproduce Low \emph{et al.}'s
$2\brot$; on their actual $2^{\brot-1}$ grid they cost $2\brot-2$; with simplified adders, $2\brot-4$; fused,
$2\brot-5$. We call the first four Toffolis per rotation, $16(N-1)$ per walk step, the accounting correction,
and the fifth, $4(N-1)$ per step, the fusion. For \FeS{} the correction lowers the step from 3\,924 to 3\,620
Toffolis and the fusion from 3\,620 to 3\,544. The combined saving is the reduction from the published costs.
Each construction is verified exhaustively at $\brot=15$, and the rotations with simplified adders and the fused rotations on the
$2^{\brot-1}$ grid at every width in use (Sec.~\ref{sec:verify}).

\begin{table}[b]
\caption{\label{tab:ladder}Exact constructions of a Givens rotation with a quantum angle on the $2^{\brot-1}$ and
$2^{\brot}$ grids: Toffolis per rotation, the value at $\brot=15$, the resulting \FeS{} walk step (76 rotations), and the
largest gadget width $b$ checked exhaustively (every smaller width was checked too); a rotation at precision $\brot$
uses $b=\brot-1$ on the $2^{\brot-1}$ grid and $b=\brot$ on the $2^{\brot}$ grid. Unless marked otherwise, rows use
Low \emph{et al.}'s $2^{\brot-1}$ grid. The rows that are not fused use two
half-angle additions. The simplified adders replace arithmetic on the two most significant gradient qubits with
Clifford phase gates. $^a$Low \emph{et al.}'s count. $^b$Also proved for every $b$ (Appendix~\ref{app:givens-proof}).
$^c$The fallback if the grid or the angle convention were one bit finer: 2 Toffolis from simplified adders and 1 from
the fusion.}
\begin{ruledtabular}
\footnotesize\setlength{\tabcolsep}{2.5pt}
\begin{tabular}{lcccc}
Construction & Toffolis & $\brot{=}15$ & \FeS{} step & Checked \\
\hline
Full adders, $2^{\brot}$ grid$^a$ & $2\brot$ & 30 & 3\,924 & $b=15$ \\
Full adders & $2\brot-2$ & 28 & 3\,772 & $b=15$ \\
Simplified adders & $2\brot-4$ & 26 & 3\,620 & $b=17$ \\
Fused & $2\brot-5$ & 25 & 3\,544 & $b=17^b$ \\
\shortstack[l]{Simplified adders,\\$2^{\brot}$ grid$^c$} & $2\brot-2$ & 28 & 3\,772 & $b=17$ \\
Fused, $2^{\brot}$ grid$^c$ & $2\brot-3$ & 27 & 3\,696 & $b=17^b$ \\
\end{tabular}
\end{ruledtabular}
\end{table}

\section{Exact verification beyond classical functions}\label{sec:verify}

\subsection{Verification conditions}\label{sec:conditions}

Gu \emph{et al.}'s evaluator follows one computational-basis trajectory at a time, which is exact for kickmix
circuits: reversible classical gates, sign gates, and $X$-basis measurements whose outcomes control only sign
corrections~\cite{Gu2026,Babbush2026kick}. The rotations above leave that model in three ways: the change of basis $F$ is
Clifford but not classical, the core uses $S$ and $S^\dagger$, and the phase-gradient register holds a fixed
resource state rather than a classical input.

We consider circuits of the form $V=F^\dagger CF$, where $F$ is a fixed unitary on some of the qubits
of a control register $c$ (a Clifford unitary in all our applications), and the core $C$ is built from $X$, CNOT, CZ, $Z$, $S$, $S^\dagger$, temporary
logical ANDs (one Toffoli each) and their measurement-based uncomputation (an $X$-basis measurement of the AND target, a
CZ on its inputs when the outcome is 1, and a reset). $C$ acts on $c$, on a $b$-qubit resource register $y$ ($b\ge2$)
prepared in $\ket{G_b}$, and on temporaries prepared in $\ket{0}$.

Each core gate maps a computational basis state to a single basis state times a power of $i$, and measurement outcomes
control only diagonal Clifford corrections. On a basis input the core therefore produces a definite output for each
measurement record, which a classical simulator can follow.

For a record $r$ of the $M$ measurements, let $V_r$ denote the corresponding branch of $V$, rescaled by $2^{M/2}$.
For an integer-valued function $T(c)$, set $D_T=\sum_ce^{2\pi iT(c)/2^b}\proj{c}$.

\begin{theorem}[Exact verification]\label{thm:verify}
For circuits of the form described above, the following conditions on every basis input $\ket{c,y,0}$,
$y\in\mathbb{Z}_{2^b}$, are necessary and sufficient for
\[
\begin{aligned}
&V_rF^\dagger\ket{c}\ket{G_b}\ket{0}\\
&\qquad=e^{2\pi iT(c)/2^b}F^\dagger\ket{c}\ket{G_b}\ket{0}
\end{aligned}
\]
to hold for every record $r$ and every basis state $c$:
(a) at every uncomputation, the measured temporary equals the AND of its two inputs;
(b) at the end, the temporaries are $\ket0$ and $c$ is unchanged;
(c) the core maps $\ket{c,y}\mapsto i^{k(c,y)}\ket{c,\pi_c(y)}$ with each $\pi_c$ a bijection of $\mathbb{Z}_{2^b}$;
(d) $\pi_c(y)-y+2^{b-2}k(c,y)\equiv T(c)\pmod{2^b}$.
By linearity, each rescaled branch acts as $F^\dagger D_TF$ on arbitrary, possibly entangled, control inputs,
returns $\ket{G_b}$ and the temporaries unchanged, and acts as the identity on external registers.
\end{theorem}

The proof is in Appendix~\ref{app:verification-proof}. The criterion does not require $F$ to be Clifford; in our
application this choice makes Eq.~(\ref{eq:frame}) a short Pauli computation (Appendix~\ref{app:givens-proof}). Conditions (c)
and (d) give the phase $e^{2\pi iT(c)/2^b}$ on $\ket{G_b}$, while (a) makes both normalized measurement branches agree.
If a record need only match up to a record-dependent global phase, (a) and (d) can be relaxed (Appendix~\ref{app:verification-proof}); a temporary computed as a NAND, for example, fails (a) on every input but changes only the sign of
some branches. The class
covers rotations $e^{i\theta P}$ of a Pauli string $P$ by a loaded angle (with $F$ mapping $P$ to $Z_1$), their controlled
versions (through mask ANDs), Givens rotations, and products of commuting rotations that share a change of basis. For a Givens
rotation, $c=(p',q',v)$ collects the system qubits in the transformed basis and the angle register, and $T(c)=(-1)^{p'\oplus\mathrm{inv}}q'(v\bmod
2^b)$, with $\mathrm{inv}=0$ for the forward and 1 for the inverse form, gives $D_T=D(\theta)$ of Eq.~(\ref{eq:frame}) or its
inverse. Appendix~\ref{app:givens-proof} checks (a)--(d) for the fused circuit at arbitrary width; the exhaustive runs below establish them independently at each width used.

\subsection{What the certificate adds}\label{sec:adds}

The four checks in Theorem~\ref{thm:verify} form a complete certificate after factoring out a fixed change of basis. First, an operation that creates superpositions is
verified by following basis states of its core. The change-of-basis identity is established once (Appendix~\ref{app:givens-proof},
checked numerically in Table~\ref{tab:verify}). At $b=14$, the fused rotation's core acts on 56 qubits with
$2^{25}$ measurement records. The check follows one classical trajectory per basis input, $2^{32}$ in all,
covering every record at once in 49\,s on five processor cores.

Second, the specification is relational, not functional. Conditions (c) and (d) say exactly that $\ket{G_b}$ is
an eigenvector of the core's phase-permutation with eigenvalue $e^{2\pi iT(c)/2^b}$; they do not prescribe the
classical map $y\mapsto\pi_c(y)$. For the circuits in Table~\ref{tab:ladder}, the full-adder rotation leaves the
gradient register in $y+T(c)$ with phase $+1$ on every basis input, but the simplified and fused rotations do so on
only 5/8 of the inputs, at every width $b=2,\dots,14$; on the other 3/8 a phase in $\mathbb{Z}_4$ compensates a
different gradient value, while (d) holds everywhere (Appendix~\ref{app:evaluator}). An evaluator that compares
outputs with a fixed reference, such as the textbook full-adder output, would reject both exact circuits unless
its reference were rewritten for each construction. Condition (d) needs no reference. The circuits' $S$ gates
also lie outside the kickmix gate set.

Third, under the branch-wise convention a failed condition is an error of the branch operators
(Theorem~\ref{thm:verify}): a circuit that passes is exact, and one that fails differs from the target
on some input and record, if only by a record-dependent global phase, as for the NAND temporary above.

\subsection{Exhaustive runs}

\begin{figure*}[t]
\includegraphics[width=\textwidth]{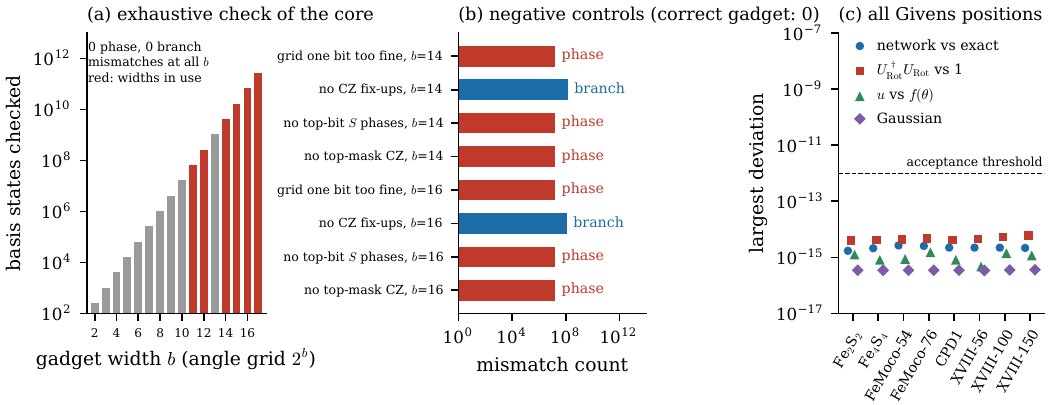}
\caption{\label{fig:verification}Verification of the fused rotation. (a)~Basis states checked exhaustively at each
width $b$, all with zero phase and zero measurement-branch mismatches; red bars are the widths the published walks use,
$b=\brot-1\in\{11,12,14,15,16,17\}$, with $2b-3=19$ to 31 Toffolis. (b)~Negative controls on one batch of $2^{24}$ basis
states at $b=14$ and $b=16$; every fault is caught, while the correct gadget gives zero. Phase mismatches count basis
states, branch mismatches count (basis state, measurement) pairs. (c)~All $4(N-1)$ Givens positions of a walk step for
each system of Table~\ref{tab:systems}, composed in the Majorana representation from the verified gadget unitaries:
largest deviation from the exact network, of $U_{\mathrm{Rot}}^\dagger U_{\mathrm{Rot}}$ from the identity, of the rotated
vector from $f(\theta)$, and of each gadget from a Gaussian unitary commuting with $Z_pZ_q$ (the check of
Eq.~(B21) of Ref.~\cite{Low2025} is exactly zero); the dashed line is the $10^{-12}$ tolerance the checks assert.}
\end{figure*}

We implemented the check in a bit-plane simulator that holds one packed plane per qubit and two planes for the phase
modulo 4, and applies each gate to $2^{24}$ basis states at once with integer operations (Appendix~\ref{app:evaluator}).
For the fused circuit it ran over every $(b+1)$-bit angle value, every gradient basis state, both values of each system
qubit, forward and inverse: $2^{2b+4}$ basis states for each width $b=2,\dots,17$, or $2.7\times10^{11}$ at $b=17$, which
took 62 minutes on seven processor cores. No width shows a phase mismatch, a measurement branch that differs, a gradient
map that is not a bijection, a dirty ancilla, a changed angle or system qubit, or a touch of the bit $v_b$, and Toffolis,
measurements and peak temporaries are exactly $2b-3$ (Fig.~\ref{fig:verification}a and Table~\ref{tab:verify}). The same
evaluator verifies the rotations with full or simplified adders of Table~\ref{tab:ladder}, two half-angle additions into a $(b+1)$-qubit
gradient register, with full adders ($2b$ Toffolis) for $b\le15$ and with simplified adders ($2b-2$) for $b\le17$. Four
deliberate faults in the fused circuit are each detected on millions of basis states: a target grid one bit too fine,
dropped CZ fix-ups, dropped top-bit $S$ phases, and a dropped top-mask CZ (Fig.~\ref{fig:verification}b). For $b\le5$ a
statevector simulator with every measurement-outcome sequence enumerated confirms the full gadget against $G(2\pi v/2^b)$
for all $v$ at once ($3.0\times10^{-15}$) and, on four outcome sequences per run, agrees with the bit-plane simulator.

Each gadget equals a Givens rotation exactly. We compose the gadgets to check the network conventions: the order
of $U_{\mathrm{Rot}}$, the inverse gadgets in $U_{\mathrm{Rot}}^\dagger$, and Low \emph{et al.}'s Eq.~(B21),
which requires $U_{\mathrm{Rot}}$ to rotate both Majorana operators of a mode by the same unit vector. Givens
rotations are fermionic Gaussian unitaries~\cite{Terhal2002,Jiang2018}, so a walk step's $4(N-1)$ rotations
compose as $2N\times2N$ orthogonal matrices. For each system, at its own $N$ and grid, the network built from
the verified gadget unitaries matches the exact network, $U_{\mathrm{Rot}}^\dagger$ undoes $U_{\mathrm{Rot}}$,
the two sets of Majorana operators are rotated identically, and the rotated vector equals the spherical-coordinate map
$u=f(\theta)$ of this gate convention (Low \emph{et al.} do not write $f$ out), all to $6\times10^{-15}$ or
better (Fig.~\ref{fig:verification}c).

\begin{table*}[t]
\caption{\label{tab:verify}Checks, forward and inverse forms except the negative controls (forward). Deviations are
maximum absolute differences of state amplitudes (statevector rows) or matrix elements (change-of-basis and network rows).}
\begin{ruledtabular}
\begin{tabular}{>{\raggedright}p{8.4cm}>{\raggedright}p{5.0cm}p{3.8cm}}
Check & Scope & Result \\
\hline
Fused rotation, statevector: gadget vs $G(2\pi v/2^b)$, all $v$ in superposition with random data, every outcome
record & $b=2$--5 & deviation $\le3.0\times10^{-15}$ \\
Fused rotation, statevector: 4-mode network $U^\dagger MU$ for $M=Z_0,X_0,Y_0$ & $b=3$ & deviation
$\le1.8\times10^{-17}$ \\
Bit-plane simulator vs statevector, same core & $b=2$--5 & deviation $\le2.8\times10^{-15}$ \\
Fused rotation, conditions (a)--(d) of Theorem~\ref{thm:verify} and counts, every basis state & $b=2$--17; $2^{2b+4}$
states each & no mismatch; $2b-3$ Toffolis \\
Rotations with full or simplified adders, same conditions, every basis state & $b=2$--15; $b=2$--17 & no
mismatch; $2b$; $2b-2$ \\
Comparison with the textbook adder (gradient register ends in $y+T$ with phase $+1$) for the three cores, every basis state &
$b=2$--14 & full adders: never violated; simplified, fused: violated on 3/8 \\
Change-of-basis identity $F^\dagger D_vF=G(2\pi v/2^b)$ for all $v$, and $2\pi$ periodicity & fused $b=11,14$--$17$; full and simplified adders: all $b$ & deviation $\le1.3\times10^{-15}$ \\
Negative controls: grid one bit too fine, no CZ fix-ups, no top-bit $S$, no top-mask CZ & $b=14$ and 16 & all detected
\\
Walk-step networks: all $4(N-1)$ positions, random and edge angles, two steps & the eight systems of
Table~\ref{tab:systems} and \FeSS{} of Table~\ref{tab:table11}, $N=20$--150 & deviation $\le6.0\times10^{-15}$ \\
\end{tabular}
\end{ruledtabular}
\end{table*}

\subsection{Scope of verification}

The verification covers the rotations, 44--72\% of each walk step; the lookups, reflections and swaps that make up the
other 28--56\% are taken from Low \emph{et al.}'s cost formulas and are neither built nor verified here. Exactness matters
because QPE applies the walk coherently $10^{4}$ to a few $10^{5}$ times. Gu \emph{et al.} validated their \FeS{} circuit, whose
input space of 72 qubits is too large to enumerate, on every table row and $2^{24}$ random inputs with sampled measurement
records~\cite{Gu2026}. By their own bounds this gives a defect probability
$\varepsilon\le\ln(1/\delta)/k\approx2.7\times10^{-7}$ at confidence $1-\delta=0.99$, and hence a state error per
application of at most $2\sqrt{\varepsilon}\approx1.0\times10^{-3}$, for inputs drawn from their test distribution rather
than the states QPE produces; it does not bound the error of $10^{5}$ coherent applications. For a gadget small enough to
enumerate, here $2b+3$ qubits besides temporaries, Theorem~\ref{thm:verify} removes that gap; extending this verification
to the rest of the walk is open.

\section{\FeS: the matched estimate and a cross-representation comparison}\label{sec:fe2s2}

Low \emph{et al.} and Gu \emph{et al.} both use the \FeS{} active space of Li and Chan, 30 electrons in 20 spatial
orbitals~\cite{LiChan2017,Sharma2014}, which we take from the authors' public repository.

\subsection{The DFTHC+BLISS+SA walk step}

With $N=20$ spatial orbitals, DFTHC rank $R=14$, $B=15$ basis vectors and $C=5$ copies per rank, and
$\lambda_{\mathrm{eff}}=6.469\Ha$, Low \emph{et al.} report 3\,924 Toffolis and 463 logical qubits per step under their
Table~V budget ($\sigma_{\mathrm{PEA}}=1\mHa$). The formulas of their released code reproduce both numbers exactly, and
uniquely within our scan (Appendix~\ref{app:walk}), with the bit precisions of their Table~X,
$(b_{\mathrm{coeff}},\brot)=(11,15)$, and block exponents $(1,3,1,3)$ for its four QROAM (select-swap) lookups. The step
contains $4(N-1)=76$ Givens rotations, 2\,280 Toffolis or 58.1\% (Fig.~\ref{fig:fe2s2}b). Using simplified adders
on the $2^{14}$ grid lowers it to 3\,620 Toffolis and the fused rotation to 3\,544, an aggregate reduction of
9.7\%, of which the fusion contributes 1.9 percentage points (76 Toffolis, or 2.1\% of the 3\,620-Toffoli step). Over the unchanged 10\,162 steps, phase estimation
then needs $3.99\times10^{7}$, $3.68\times10^{7}$ and $3.60\times10^{7}$ Toffolis. In the surface-code model Low
\emph{et al.} use (Appendix~\ref{app:cost}), four AutoCCZ(15,23) magic-state factories~\cite{GidneyFowler2019} at code
distance 23 run these circuits in 0.318, 0.294 and 0.288\,h on the same 1.35 million physical qubits; the failure
probability falls from 0.086 to 0.077. Under their statistical budget (their Table~XI, $\sigma_{\mathrm{PEA}}=1.4\mHa$, a
different factorization with $\brot=16$) the cost falls from $3.33\times10^{7}$ to $3.12\times10^{7}$ and
$3.06\times10^{7}$ Toffolis (Table~\ref{tab:table11}).

\subsection{Comparison with the sparse Pauli walk}\label{sec:sparse}

Gu \emph{et al.} map the 40 spin-orbitals to 40 qubits and keep Pauli terms with $|w_\ell|\ge10^{-4}\Ha$, sampled by an
alias table~\cite{Walker1977}. Expanding the Hamiltonian in Majorana monomials, each of which is one Pauli string for any
Jordan--Wigner ordering, we find 233\,000 nonzero terms and exactly their $L=41\,052$ after truncation (796 with two
Majorana operators, 40\,256 with four; Appendix~\ref{app:sparse}). Gu \emph{et al.} do not state the normalization; we
find $\lambda=\sum_\ell|w_\ell|=133.465\Ha$ ($137.260\Ha$ without truncation), so QPE to $\sigma_{\mathrm{PEA}}=1\mHa$
needs $\lceil\pi\lambda/2\sigma_{\mathrm{PEA}}\rceil=209\,647$ walk steps. Their optimized circuit is reported as
$1.5\times10^{4}$ Toffolis and 126 qubits; its components sum to 14\,978 Toffolis. It is only the data-dependent part of a
step, the alias lookup, the selected Pauli operator and its erasure, without the uniform-state preparations and the
reflection. Multiplied by the step count it gives $3.14\times10^{9}$ Toffolis, and at least $3.04\times10^{9}$ for any
count that rounds to $1.5\times10^{4}$; the same estimate for their baseline is $1.15\times10^{10}$.

\begin{figure*}[t]
\includegraphics[width=\textwidth]{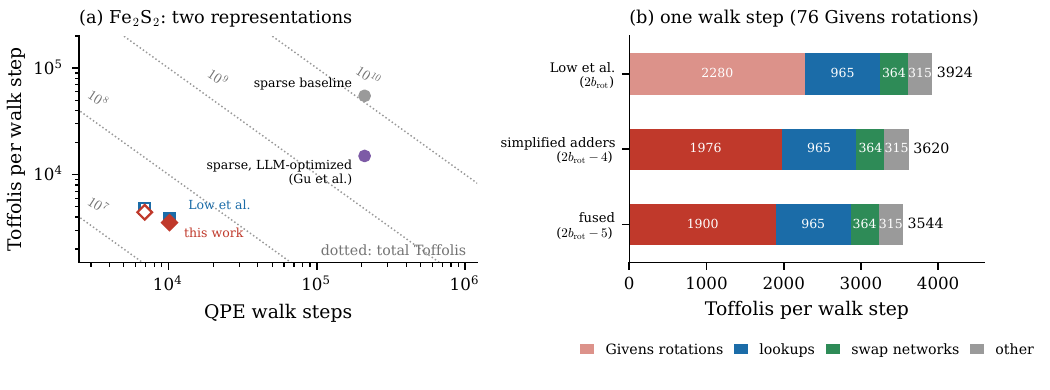}
\caption{\label{fig:fe2s2}\FeS{} [30e, 20o]. (a)~The two factors of the QPE cost, walk steps and Toffolis per step, for
the sparse Pauli walk of Gu \emph{et al.} (data-dependent part only) and the DFTHC+BLISS+SA walk of Low \emph{et al.}
without and with the fused rotations; dotted lines are constant total Toffolis. Filled markers: $\sigma_{\mathrm{PEA}}=1\mHa$;
open markers: Low \emph{et al.}'s statistical budget, $\sigma_{\mathrm{PEA}}=1.4\mHa$. The two representations differ in
their approximations (Sec.~\ref{sec:sparse}); the circuit-level changes move a point vertically. (b)~One DFTHC+BLISS+SA walk
step rebuilt from Low \emph{et al.}'s formulas (their Table~V budget) with their rotation cost, with simplified adders on their grid, and with the fused rotations; only the 76 Givens rotations change.}
\end{figure*}

These resource estimates use different representations and approximations, so they do not give a matched
comparison of circuits: the $10^{-4}\Ha$ cut discards $3.8\Ha$ of 1-norm, whose effect on the energy is not
assessed, while Low \emph{et al.} estimate their factorization error at 0.013\mHa; a tighter cut would raise
$\lambda$ and the sparse cost. The sparse figure omits part of the walk step, so it understates that walk's
cost, and we did not try to improve Gu \emph{et al.}'s circuit. In Table~\ref{tab:fe2s2} and
Fig.~\ref{fig:fe2s2}a, a DFTHC+BLISS+SA step costs fewer Toffolis than the data-dependent part alone of the
optimized sparse step, and QPE needs 20.6 times fewer steps, so the published DFTHC+BLISS+SA estimate is already
lower than the sparse one by a factor of 79. Our rotations give a further reduction by a factor of 1.11, for an
overall factor of about 87 (at least 84, given the rounding of Gu \emph{et al.}'s count). Of the factor of 20.6
in walk steps, 7.6 comes from the smaller normalization of DFTHC with BLISS, $\Lambda=17.53\Ha$ against
$\lambda=133.47\Ha$, and 2.7 from spectral amplification. Their combined effect is larger than the circuit-level
reductions within one walk: factors of 3.7 for the sparse subroutine of Ref.~\cite{Gu2026} and 1.11 here.

\begin{table*}[t]
\caption{\label{tab:fe2s2}Phase estimation for \FeS{} [30e, 20o] to $\sigma_{\mathrm{PEA}}=1\mHa$ from the same integrals.
The sparse and DFTHC+BLISS+SA rows are stated resource estimates under different representations and approximation
errors, not a matched comparison (Sec.~\ref{sec:sparse}). Sparse rows: Gu \emph{et al.}'s per-step Toffolis for the
data-dependent part only (component sum 14\,978 for the optimized circuit, reported 55\,000 for the baseline) and qubits,
times the step count from our $\lambda$. DFTHC+BLISS+SA rows: Low \emph{et al.}'s published step, the same with simplified adders on their angle grid, and with the fused rotations; these three are matched. Run times and
physical qubits: the surface-code model of Low \emph{et al.} with four CCZ factories and their layout rule
(Appendix~\ref{app:cost}); for the sparse rows they follow from the reported qubit counts and are model estimates, not
bounds. $TQ$: Toffolis times logical qubits, Gu \emph{et al.}'s figure of merit, here for the whole estimation.}
\begin{ruledtabular}
\footnotesize
\begin{tabular}{lcccccccc}
Route & $\lambda$ (Ha) & Steps & Toffolis/step & Qubits & QPE Toffolis & QPE $TQ$ & Run time & Phys.\ qubits\\
\hline
Sparse walk, baseline~\cite{Gu2026} & 133.465 & 209\,647 & 55\,000 & 139 & $1.15\times10^{10}$ & $1.6\times10^{12}$ & 108\,h & 1.13\,M \\
Sparse walk, LLM-optimized~\cite{Gu2026} & 133.465 & 209\,647 & 14\,978 & 126 & $3.14\times10^{9}$ & $4.0\times10^{11}$ & 29.4\,h & 1.05\,M \\
DFTHC+BLISS+SA~\cite{Low2025} & 6.469 & 10\,162 & 3\,924 & 463 & $3.99\times10^{7}$ & $1.85\times10^{10}$ & 0.318\,h & 1.35\,M \\
\quad + simplified adders & 6.469 & 10\,162 & 3\,620 & 463 & $3.68\times10^{7}$ & $1.70\times10^{10}$ & 0.294\,h & 1.35\,M \\
\quad + fused rotations (this work) & 6.469 & 10\,162 & 3\,544 & 463 & $3.60\times10^{7}$ & $1.67\times10^{10}$ & 0.288\,h & 1.35\,M \\
\end{tabular}
\end{ruledtabular}
\end{table*}

\section{FeMoco and the other benchmark systems}\label{sec:systems}

Low \emph{et al.} report eight systems under the error budget of their Table~V: \FeS{} and \FeSS{} clusters~\cite{LiChan2017}, FeMoco
with 54 and 76 orbitals~\cite{Reiher2017,Li2019}, a model of the compound~I intermediate of cytochrome P450
(CPD1)~\cite{Goings2022}, and a ruthenium CO$_2$-fixation catalyst (XVIII) with 56, 100 and 150
orbitals~\cite{vonBurg2021}. For every one of them, a scan over the bit precisions and QROAM block sizes finds exactly one
parameter set for which the formulas of their released code give their published Toffolis and logical qubits per step,
and it uses the bit precisions of their Tables~X and~XII (Table~\ref{tab:systems}). Their $\brot$ values are 12, 15, 16 and
17, so the fused rotation runs at widths 11, 14, 15 and 16, all verified exhaustively. A step has $4(N-1)$ rotations, so it
saves $20(N-1)$ Toffolis whatever $\brot$ is. The aggregate saving, 9.5--14.3\%, is largest where Givens rotations
dominate the step and $\brot$ is small; the fusion accounts for 1.9--2.9 percentage points of it.

\begin{table*}[t]
\caption{\label{tab:systems}The eight benchmark systems of Low \emph{et al.}~\cite{Low2025} under their Table~V error budget,
$\sigma_{\mathrm{PEA}}=1\mHa$ with $\lceil\pi\lambda_{\mathrm{eff}}/2\sigma_{\mathrm{PEA}}\rceil$ steps, unchanged. For
each system a scan over $\brot\in[10,20]$, $b_{\mathrm{coeff}}\in[5,16]$ and four QROAM exponents in $[0,7]$ finds exactly
one parameter set for which the formulas of their released code (Appendix~\ref{app:walk}) give their published Toffolis
and logical qubits per step, and it has their stated $(b_{\mathrm{coeff}},\brot)$. Saving: aggregate, relative to the
published step, and in parentheses the fusion's part, in percentage points. Run times:
four AutoCCZ$(\ell_1,\ell_2)$ factories, with the factory type and code distance $d$ that the model chooses for Low
\emph{et al.}'s circuit kept for ours (Appendix~\ref{app:cost}); physical qubits are the same for both.}
\begin{ruledtabular}
\footnotesize\setlength{\tabcolsep}{2.2pt}
\begin{tabular}{lccccccccc}
System & $N$ & $(b_{\mathrm{coeff}},\brot)$ & Givens & Toffolis per step & Saving (\%) & QPE Toffolis & $(\ell_1,\ell_2)$, $d$ & Run time & Phys.\ qubits \\
\hline
\FeS{} [30e, 20o] & 20 & (11, 15) & 58.1\% & 3\,924 $\to$ 3\,544 & 9.7 (1.9) & $3.99\to3.60\times10^{7}$ & (15,23), 23 & 0.318 $\to$ 0.288\,h & 1.35\,M \\
\FeSS{} [54e, 36o] & 36 & (13, 17) & 64.7\% & 7\,356 $\to$ 6\,656 & 9.5 (1.9) & $1.73\to1.57\times10^{8}$ & (15,23), 25 & 1.38 $\to$ 1.25\,h & 2.31\,M \\
FeMoco [54e, 54o] & 54 & (9, 16) & 66.5\% & 10\,203 $\to$ 9\,143 & 10.4 (2.1) & $3.42\to3.07\times10^{8}$ & (15,23), 27 & 2.73 $\to$ 2.45\,h & 3.22\,M \\
FeMoco [113e, 76o] & 76 & (9, 15) & 61.5\% & 14\,629 $\to$ 13\,129 & 10.3 (2.1) & $1.00\to0.90\times10^{9}$ & (17,25), 27 & 8.71 $\to$ 7.81\,h & 4.37\,M \\
CPD1 [63e, 58o] & 58 & (10, 15) & 71.7\% & 9\,535 $\to$ 8\,395 & 12.0 (2.4) & $4.91\to4.32\times10^{8}$ & (15,23), 27 & 3.92 $\to$ 3.45\,h & 3.26\,M \\
XVIII [64e, 56o] & 56 & (7, 12) & 68.7\% & 7\,685 $\to$ 6\,585 & 14.3 (2.9) & $2.06\to1.77\times10^{8}$ & (15,23), 27 & 1.65 $\to$ 1.41\,h & 2.72\,M \\
XVIII [100e, 100o] & 100 & (8, 16) & 70.2\% & 18\,041 $\to$ 16\,061 & 11.0 (2.2) & $1.07\to0.95\times10^{9}$ & (15,25), 29 & 9.27 $\to$ 8.25\,h & 5.89\,M \\
XVIII [150e, 150o] & 150 & (9, 16) & 69.7\% & 27\,367 $\to$ 24\,387 & 10.9 (2.2) & $2.83\to2.52\times10^{9}$ & (17,25), 29 & 24.6 $\to$ 21.9\,h & 8.68\,M \\
\end{tabular}
\end{ruledtabular}
\end{table*}

\begin{figure*}[t]
\includegraphics[width=\textwidth]{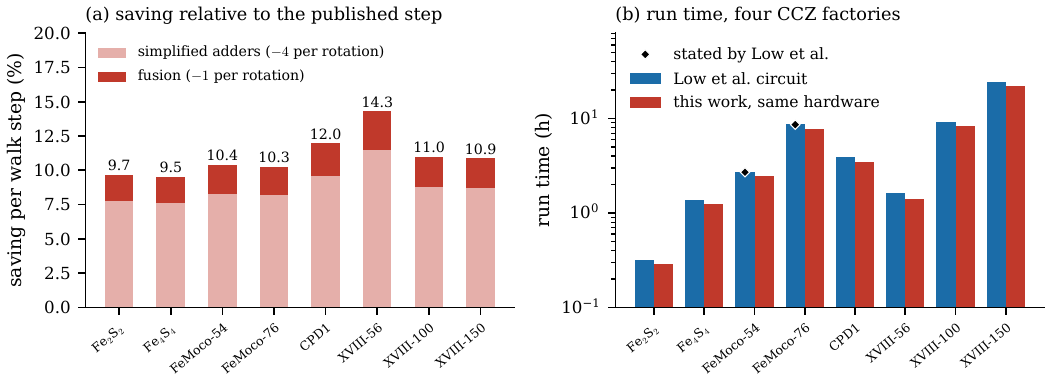}
\caption{\label{fig:systems}All eight systems of Low \emph{et al.} under their Table~V error budget. (a)~Saving per walk step
relative to the published step, split into the accounting correction (simplified adders on their angle grid, 4 Toffolis per rotation) and the fusion (1 Toffoli per rotation); the number on each bar is the aggregate.
(b)~Run time of phase estimation in our transcription of their surface-code model with four CCZ factories, on the factory
type and code distance chosen for Low \emph{et al.}'s circuit; diamonds are the run times Low \emph{et al.} state.}
\end{figure*}

Our transcription of the surface-code model agrees to within 3\% with the two physical estimates that Low \emph{et al.}
state: 8.71\,h on 4.37 million physical qubits at $d=27$ for FeMoco-76 (they give 8.6\,h and 4.5 million), and 2.73\,h on
3.22 million at $d=27$ for FeMoco-54 (they give 2.7\,h and 3.2 million). On the same factories and code distances the run
times fall in proportion to the Toffoli counts: for FeMoco-76 from 8.71\,h to 7.99\,h with simplified adders
and to 7.81\,h with the fused rotations, and for the largest catalyst model from 24.6\,h to 21.9\,h
(Fig.~\ref{fig:systems}b). The failure probability falls with the run time, from 0.079 to 0.071 for FeMoco-76.

Low \emph{et al.}'s Table~XI gives a second set of estimates for six of the systems, with an error budget that combines the
factorization and rounding errors statistically and allows $\sigma_{\mathrm{PEA}}=1.4\mHa$; they caution that it is not
directly comparable to prior work. The formulas in their released code reproduce the costs of all six walk steps exactly (Appendix~\ref{app:walk}).
The saving there is 6.1--12.9\% (Table~\ref{tab:table11}), of which the fusion accounts for 1.2--2.6 percentage points;
the width $b=17$ that \FeSS{} needs was verified separately, for the fused rotation and the rotation with simplified adders, on $2^{38}$ basis
states each. For
\FeS{} this budget gives the lowest cost in either convention, $3.06\times10^{7}$ Toffolis.

\begin{table}[t]
\caption{\label{tab:table11}Low \emph{et al.}'s statistical budget (their Table~XI, $\sigma_{\mathrm{PEA}}=1.4\mHa$), all
six walk steps rebuilt exactly. Saving as in Table~\ref{tab:systems}; QPE Toffolis without and with the fused rotations.}
\begin{ruledtabular}
\footnotesize\setlength{\tabcolsep}{2pt}
\begin{tabular}{lcccc}
System & $\brot$ & Toffolis per step & Saving (\%) & QPE Toffolis \\
\hline
\FeS{} & 16 & 4\,814 $\to$ 4\,434 & 7.9 (1.6) & $3.33\to3.06\times10^{7}$ \\
\FeSS{} & 18 & 11\,493 $\to$ 10\,793 & 6.1 (1.2) & $1.81\to1.70\times10^{8}$ \\
FeMoco-54 & 17 & 12\,930 $\to$ 11\,870 & 8.2 (1.6) & $3.11\to2.86\times10^{8}$ \\
FeMoco-76 & 16 & 17\,588 $\to$ 16\,088 & 8.5 (1.7) & $1.00\to0.92\times10^{9}$ \\
CPD1 & 17 & 12\,897 $\to$ 11\,757 & 8.8 (1.8) & $4.82\to4.39\times10^{8}$ \\
XVIII-56 & 13 & 8\,554 $\to$ 7\,454 & 12.9 (2.6) & $1.72\to1.50\times10^{8}$ \\
\end{tabular}
\end{ruledtabular}
\end{table}

We keep Low \emph{et al.}'s logical qubit counts throughout. The fused rotation needs $2\brot-5$ temporaries at
peak, 19 to 31; their Table~IX budgets $\brot$ temporaries for the additions, and their released code none for
SEL. Low \emph{et al.} size temporary space by the largest temporary requirement of a step, which for these
systems is the outer alias lookup's 47 to 106 qubits, released before the rotations run, and the rotation's
temporaries fit in it. Counting them separately instead would add $\brot-5=7$ to 13 logical qubits relative to
Table~IX (at most 2.2\%, for \FeS{}) or $2\brot-5$ relative to the code (at most 5.4\%), and would change no run
time in Table~\ref{tab:systems}. These estimates do not include qubit savings from the angle register, which
needs one bit fewer per angle, or the gradient register, which needs one qubit fewer.

\section{Discussion}\label{sec:discussion}

Theorem~\ref{thm:verify} gives exact, complete verification when the non-classical parts of a circuit consist
of a fixed change of basis (Clifford in all our applications), phases in $\mathbb{Z}_4$, and a resource state that is an
eigenstate of the core's permutations. This structure recurs in fault-tolerant algorithms: in rotations by
loaded angles, in the basis rotations of double-factorized and tensor-hypercontracted
walks~\cite{vonBurg2021,Lee2021,Low2025,Caesura2025}, and in the controlled rotations of state-preparation
trees. Gates whose change of basis is itself controlled, such as controlled Majorana selections, would need an extension
of the theorem.

The change in the published costs is mostly a correction of their accounting, and the fused circuit contributes
about 2\% of a walk step. The estimates depend on the following assumptions. (1)~The baseline is the $2\brot$
per rotation of Low \emph{et al.}'s totals and released code (Sec.~\ref{sec:cost}). (2)~The 25-Toffoli rotation
relies on two readings of Low \emph{et al.}'s appendix: the $2^{\brot-1}$ grid and the Majorana-rotation
convention for $\theta$. If either fails, the angle lives on a $2^{\brot}$ grid, on which our negative control
shows the circuit would be wrong; the same fused construction at $b=\brot$ then applies, at $2\brot-3$ Toffolis,
and the saving per step falls from $20(N-1)$ to $12(N-1)$, $8(N-1)$ from simplified adders and $4(N-1)$ from
the fusion, or 5.7--8.6\% under Low \emph{et al.}'s Table~V budget (6.2\% for FeMoco-76, 5.8\% for \FeS{}). (3)~The
surface-code model follows Low \emph{et al.} in treating Clifford gates and measurements as free; our rotation
uses $2\brot-5$ measurements with classically controlled CZ per rotation, and at four factories each Toffoli
takes at least $28.75\,\mu$s, well above the $10\,\mu$s reaction time, so feed-forward never binds. (4)~As in
Low \emph{et al.}, the trial state is assumed to have unit ground-state overlap and its
preparation~\cite{Berry2025} is not costed.

The rotation savings can be combined with more factories, concurrent Toffolis or a tighter sum-of-squares gap,
all of which leave the rotation network unchanged. The evaluator also offers a possible fitness function for
circuit search: it is a few hundred lines of code and checks $2^{36}$ basis states in 14 minutes on five
processor cores. We have not run such a search. Applying an LLM-guided search such as that of Gu \emph{et
al.}~\cite{Gu2026} to the complete DFTHC+BLISS+SA walk step, using this evaluator for its non-classical parts
and the kickmix evaluator for the rest, would test whether exact verification of quantum-angle rotations can
lead to further circuit improvements.

\section{Conclusion}\label{sec:conclusion}

Exhaustive basis-state testing verifies not only classical reversible circuits but every circuit of the form described in
Sec.~\ref{sec:conditions} on phase-gradient resource states. The four verification conditions are sufficient and,
under the branch-wise convention, necessary. This class includes the Givens rotations with quantum angles that
account for 44--72\% of the Toffolis in Low \emph{et al.}'s walks. We verify each construction on the angle
grid that, on our reading, their walk already uses. A rotation with simplified adders costs $2\brot-4$ Toffolis per rotation
and a fused masked addition costs $2\brot-5$, compared with $2\brot$ in their totals and released code.

We reproduce the published costs of all 14 walk steps exactly. Replacing the rotations lowers the
phase-estimation cost by 9.5--14.3\% under Low \emph{et al.}'s Table~V budget and 6.1--12.9\% under the statistical budget,
with the walk operator unchanged. For \FeS{}, the resulting cost is $3.60\times10^{7}$ Toffolis
($3.06\times10^{7}$ under the statistical budget). For FeMoco-76, our transcription of the surface-code model
gives 7.81\,h instead of 8.71\,h on the same hardware.

\section*{Acknowledgments}
Claude (Anthropic) assisted with circuit development, verification and analysis code, and manuscript drafting.
OpenAI Codex assisted with subsequent manuscript revision. The authors directed the work and checked the results
using the verification procedures and analysis scripts described in this paper. The authors take responsibility
for the content of the manuscript.

\section*{Data availability}
The \FeS{} integrals of Li and Chan are available at
\url{https://github.com/zhendongli2008/Active-space-model-for-Iron-Sulfur-Clusters}, and the resource-estimation code
of Low \emph{et al.} is available in Ref.~\cite{Low2025code}. The circuit implementations, verification and analysis
scripts, and numerical results supporting this work are available from the authors on request.

\appendix

\section{Verification and circuit analysis}\label{app:proofs}

\subsection{Verification criterion}\label{app:verification-proof}

We use the notation of Theorem~\ref{thm:verify}, with $b\ge2$ and the resource state
\[
\ket{G_b}=2^{-b/2}\sum_{y=0}^{2^b-1}e^{-2\pi iy/2^b}\ket{y}.
\]

\begin{proof}[Proof of Theorem~\ref{thm:verify}]
\emph{Sufficiency.} At each measurement, write the state as $\sum_z\alpha_z\ket{z_{\neg t}}\ket{t(z)}$.
Projection onto $\ket{\pm}$ contributes $2^{-1/2}(\pm1)^{t(z)}$ to each term. By (a), $t(z)=a(z)\wedge b'(z)$,
so the CZ correction cancels the sign for outcome $-$. After resetting $t$, both branches equal the coherently
uncomputed state times $2^{-1/2}$ and have probability $1/2$. Induction over the measurements shows that every
rescaled branch follows the outcome-0 map used by the evaluator.

By (b) and (c), that map sends $\ket{c,y,0}$ to $i^{k(c,y)}\ket{c,\pi_c(y),0}$. Condition (d) gives
\[
i^{k(c,y)}e^{-2\pi iy/2^b}
=e^{-2\pi i(\pi_c(y)-T(c))/2^b}.
\]
Relabelling $y'=\pi_c(y)$ in the resource-state sum therefore gives
$C_r\ket{c}\ket{G_b}\ket{0}=e^{2\pi iT(c)/2^b}\ket{c}\ket{G_b}\ket{0}$ for every rescaled core branch $C_r$.
Conjugation by $F$ yields the claimed action of $V_r$; linearity extends it to arbitrary inputs, including those
entangled with other registers.

\emph{Necessity.} On a basis input $z=(c,y,0)$ the core follows the same basis-state trajectory for every record:
measurement outcomes affect only phases, and each measured temporary is reset. Let $e_j(z)$ be the value of
$t\oplus(a\wedge b')$ at measurement $j$. Projection onto $\ket{-}$ and the CZ correction together contribute
$(-1)^{e_j(z)}$ relative to outcome $+$, so
\[
C_r\ket{z}=(-1)^{r\cdot e(z)}i^{k(z)}\ket{w(z)},
\]
where $w$ and $k$ are independent of $r$. For each fixed $c$, the target state has $2^b$ nonzero amplitudes on the
states $\ket{c,y',0}$. The input resource state has exactly $2^b$ terms; if $y\mapsto w(c,y)$ were not injective,
the output would have fewer than $2^b$ nonzero amplitudes. Thus $w(c,y)=(c,\pi_c(y),0)$ for a bijection $\pi_c$,
which gives (b) and (c). Comparing amplitudes for $r=0$ gives
$i^{k(c,y)}e^{-2\pi iy/2^b}=e^{2\pi i(T(c)-\pi_c(y))/2^b}$, which is (d). Comparing with a record whose only
outcome 1 is at measurement $j$ gives $(-1)^{e_j(c,y)}=1$, which is (a).
\end{proof}

If equality is required only up to a record-dependent global phase, (b) and (c) remain unchanged, while (a) allows
each $e_j(z)$ to be an input-independent bit $e_j$ and (d) allows an input-independent additive constant $t_0$.
The same argument then gives $V_r=(-1)^{r\cdot e}e^{2\pi it_0/2^b}F^\dagger D_TF$.
Our evaluator uses the stricter branch-wise conditions, which all circuits in this paper satisfy.

\subsection{Givens rotations and the fused circuit}\label{app:givens-proof}

For adjacent Jordan--Wigner qubits $p,q$, let $K=X_pY_q-Y_pX_q$, so $G(\theta)=e^{-i\theta K/2}$ as in
Eq.~(\ref{eq:givens}). Conjugation by $\CNOT_{p\to q}$ sends $K$ to $Y_p(Z_q-1)$; the subsequent $S_p^\dagger$
and $H_p$ send $Y_p$ to $Z_p$. Thus
\[
FKF^\dagger=Z_p(Z_q-1)=-2Z_p\proj{1}_q,
\]
which gives Eq.~(\ref{eq:frame}) for every real $\theta$.
Also, $K$ vanishes on $\ket{00},\ket{11}$ and acts as $-2\sigma_y$ on
$\mathrm{span}\{\ket{01},\ket{10}\}$. Hence $G(\theta)=1\oplus e^{i\theta\sigma_y}$ in these two subspaces,
so $G(\theta+2\pi)=G(\theta)$ and $G(\theta+\pi)=Z_pZ_qG(\theta)$.
This is the Majorana-angle convention used in all our checks and read from Low \emph{et al.}'s App.~D.
The angle register holds $v=\sum_{i=0}^{b}v_i2^i$.

For the circuit in Fig.~\ref{fig:gadget}b (listed in Appendix~\ref{app:listing}), take $b\ge2$ and
$v\in[0,2^{b+1})$. In the transformed basis, set $c=(p',q',v)$, $s=(-1)^{p'\oplus\mathrm{inv}}$, and
$T(c)=s\,q'(v\bmod2^b)$. We check conditions (a)--(d) of Theorem~\ref{thm:verify} directly.
The masks are $m_i=q\wedge v_i$
for $i\le b-2$, computed by $b-1$ ANDs and erased at the end; $q$ and $v$ are never written, so (a) holds for the masks,
and $M=\sum_{i\le b-2}m_i2^i=q\,(v\bmod 2^{b-1})$. The signed addition complements every gradient qubit by CNOTs from $p$
(and an $X$ in the inverse form) exactly when $s=-1$, runs Gidney's ripple-carry adder~\cite{Gidney2018} on the low $b-2$
bits, and complements again. The adder computes the carries $c_{i+1}=\mathrm{MAJ}(m_i,y_i,c_i)$ for $i=0,\dots,b-3$ as the
AND of $m_i\oplus c_i$ and $y_i\oplus c_i$, writes the sum bits while it erases the carries in reverse order, and each
carry equals the AND of its restored inputs when it is measured, so (a) holds for the carries. The two top gradient qubits
are only complemented and restored.

On a basis state, let $M_{\mathrm{low}}=M\bmod2^{b-2}$ and let $c=c_{b-2}$ be the carry out of the low $b-2$ bits. For
$s=+1$ the low bits become $y_{\mathrm{low}}+M_{\mathrm{low}}-2^{b-2}c$, so $y'-y=M_{\mathrm{low}}-2^{b-2}c$; for $s=-1$
the adder acts on the complement $\bar y$, $c$ is the carry of $\bar y_{\mathrm{low}}+M_{\mathrm{low}}$, and complementing
back gives $y'-y=-(M_{\mathrm{low}}-2^{b-2}c)$. Either way $y'-y=s(M_{\mathrm{low}}-2^{b-2}c)$, and $y\mapsto y'$ is a
bijection because the adder is reversible, which is (c). The phase gates are $S^{s}$ on $m_{b-2}$ and on $c$ (step 5 of
Appendix~\ref{app:listing}) and $\mathrm{CZ}(q,v_{b-1})$, so $k\equiv s(m_{b-2}+c)+2qv_{b-1}\equiv s(m_{b-2}+c+2qv_{b-1})
\pmod4$. Hence
\begin{align*}
y'-y+2^{b-2}k&\equiv s\bigl(M_{\mathrm{low}}+2^{b-2}m_{b-2}+2^{b-1}qv_{b-1}\bigr)\\
&=s\,q\,(v\bmod2^b)\pmod{2^b},
\end{align*}
which is (d). Condition (b) holds because every temporary is erased and the system and angle qubits are only read.
Theorem~\ref{thm:verify} therefore gives the action $D_T$ in the transformed basis, with $\ket{G_b}$ and all temporaries
restored. Equation~(\ref{eq:frame}) and the $2\pi$ periodicity above then give $G(2\pi v/2^b)$ in the forward form and
$G(2\pi v/2^b)^\dagger$ in the inverse form. The bit $v_b$ is never read.
The $b-1$ mask ANDs and $b-2$ carry ANDs require $2b-3$ Toffolis and $2b-3$ measurements. All masks and carries are
alive at the top of the adder, so the peak temporary count is also $2b-3$.
For $b=2$ there are no carries, $M_{\mathrm{low}}=0$, $y'=y$, and $k\equiv s(m_0+2qv_1)$, giving the same result.

Under the angle-grid convention of Sec.~\ref{sec:givens}, substituting this circuit with $b=\brot-1$ for each of
Low \emph{et al.}'s $4(N-1)$ rotations leaves the walk unchanged and saves $5$ Toffolis per rotation,
or $20(N-1)$ per step.

\subsection{Restriction on a single addition}\label{app:oneadd-proof}

Let $b\ge2$ and $\theta_j=2\pi j/2^b$. Let $P$ be a Hermitian operator with $P^2=1$ on $p,q$ and $a$ ancilla qubits,
$\kappa$ a real number, $\ket{\alpha},\ket{\beta}$ fixed ancilla states, and $W_1,W_2$ linear maps that do not depend on
$\theta$ (for instance the branch operators of a circuit with measurements). There is no function $c$ without zeros such
that $\bra{\beta}W_2e^{i\kappa\theta_jP}W_1\ket{\alpha}=c(\theta_j)\,G(\theta_j)$ for $j=0,\dots,2^b-1$.

Since $e^{i\kappa\theta P}=\cos(\kappa\theta)+i\sin(\kappa\theta)P$, the left side is $\cos(\kappa\theta)A+\sin(\kappa
\theta)A'$ with $A,A'$ independent of $\theta$, and so is its conjugate by $F$. By Eq.~(\ref{eq:frame}), $FG(\theta)F^\dagger
=D(\theta)$, whose matrix elements in the computational basis of $(q,p)$ include $1$, $e^{i\theta}$ and $e^{-i\theta}$.
Equality would put the three functions $j\mapsto c(\theta_j)$, $c(\theta_j)e^{i\theta_j}$, $c(\theta_j)e^{-i\theta_j}$ in
the span of $j\mapsto\cos(\kappa\theta_j),\sin(\kappa\theta_j)$, which has dimension at most two. Since $c$ has no zeros, a
linear relation among them is one among $1,e^{i\theta_j},e^{-i\theta_j}$, which are distinct characters of
$\mathbb{Z}_{2^b}$ for $2^b\ge4$ and hence linearly independent; a contradiction.

A signed addition of the angle register into $\ket{G_b}$, with the sign set by the parity of a set $S$ of qubits through
CNOTs onto the gradient register (as in Fig.~\ref{fig:gadget}b), is of this form when no other gate reads the angle
register or acts on the gradient register: on each angle value it applies $e^{i\kappa\theta Z_S}$, conjugated by the
surrounding gates, and leaves $\ket{G_b}$ in place. This restriction does not apply to circuits in which other gates read the
angle register, Clifford or not, such as the mask ANDs and the $\mathrm{CZ}(q,v_{b-1})$ in Fig.~\ref{fig:gadget}b, or
alter the resource state, for instance by swapping $\ket{G_b}$ with a uniform superposition under control of $q$.

\section{Gate-by-gate listing}\label{app:listing}

With $p,q$ the system qubits, $v_0,\dots,v_{b}$ the angle register, $y_0,\dots,y_{b-1}$ the gradient register (least
significant first) and $\mathrm{inv}\in\{0,1\}$:
\begin{enumerate}
\item Change of basis: $\CNOT_{p\to q}$, $S^\dagger_p$, $H_p$.
\item Masks: for $i=0,\dots,b-2$, $m_i\leftarrow\mathrm{AND}(q,v_i)$ into a fresh temporary.
\item Sign: for every $j<b$, $\CNOT_{p\to y_j}$, and $X_{y_j}$ if $\mathrm{inv}=1$.
\item Carries (for $b\ge3$): $c_1\leftarrow\mathrm{AND}(m_0,y_0)$; for $i=1,\dots,b-3$: $\CNOT_{c_i\to m_i}$, $\CNOT_{c_i\to y_i}$,
$c_{i+1}\leftarrow\mathrm{AND}(m_i,y_i)$, $\CNOT_{c_i\to c_{i+1}}$, $\CNOT_{c_i\to m_i}$, $\CNOT_{c_i\to y_i}$.
\item Top two bits: $S$ on $c_{b-2}$ and on $m_{b-2}$, each followed by $\mathrm{CZ}(p,\cdot)$, and by $Z$ if
$\mathrm{inv}=1$ (this gives $S^{\pm1}$ with the sign of the addition); for $b=2$ only the $m_0$ term.
\item Erase carries and write sums (for $b\ge3$; nothing for $b=2$): for $i=b-3,\dots,0$: for $i\ge1$ apply $\CNOT_{c_i\to m_i}$, $\CNOT_{c_i\to y_i}$,
$\CNOT_{c_i\to c_{i+1}}$; erase $c_{i+1}$ by an $X$ measurement with a CZ fix-up on the two AND inputs, which at this
point hold $m_i\oplus c_i$ and $y_i\oplus c_i$; for $i\ge1$ undo the two CNOTs on $m_i$ and $y_i$; then
$\CNOT_{m_i\to y_i}$ and, for $i\ge1$, $\CNOT_{c_i\to y_i}$.
\item Undo the sign step 3.
\item Top mask bit: $\mathrm{CZ}(q,v_{b-1})$.
\item Erase masks: for $i=b-2,\dots,0$, $X$ measurement of $m_i$ with $\mathrm{CZ}(q,v_i)$ fix-up.
\item Inverse change of basis: $H_p$, $S_p$, $\CNOT_{p\to q}$.
\end{enumerate}
Toffolis: $(b-1)+(b-2)=2b-3$. The qubit $v_b$ appears nowhere. The rotations with full or simplified adders of Table~\ref{tab:ladder}
replace steps 2--9 by two signed additions of the $b$-bit angle into a $(b+1)$-qubit gradient register, one with its sign
set by $z_p$ and one by $z_p\oplus z_q$, each with $b$ carries (full adder) or $b-1$ carries and Clifford phase gates
on the two most significant gradient qubits (simplified adder).

\section{Evaluator}\label{app:evaluator}

For a core with control register $c=(p',q',v)$ and gradient register $y$, the evaluator runs all $2^{2b+1}$ pairs
$(v,y)$ for each $(p',q')$ and each direction, in batches of $2^{24}$ basis states with one packed bit plane per qubit
and two planes for the phase modulo 4. It checks conditions (a)--(d) of Theorem~\ref{thm:verify}, that $v_b$ is never
acted on, and the counts of ANDs, measurements and live temporaries. Because (a) makes both outcomes of each measurement
give the same state, the simulator follows the outcome-0 trajectory and evaluates (a) on it, which by induction over the
measurements covers every record. The change-of-basis identity is also checked numerically for every $v$, with the phases taken from
the run. For $b\le5$ the bit-plane simulator was cross-checked against a statevector simulator running the same core with
explicit measurements (four outcome sequences each, including all-0 and all-1), and that statevector simulator checks the
full gadget against $G$ with every outcome sequence; four negative controls test the evaluator itself.
These controls exercise failures of (a) and (d); none targets (b) or (c). The network checks
use the $2N\times2N$ Majorana transformation $R_{ac}=\tfrac14\mathrm{tr}(\gamma_c U^\dagger\gamma_a U)$ of each gadget
unitary on its two modes, with a residual that measures any non-Gaussian part. The exhaustive runs of the fused circuit
take 49\,s ($b=14$), 195\,s ($b=15$) and 833\,s ($b=16$) on five cores and 3\,727\,s ($b=17$) on seven; those of the
rotations with full or simplified adders take 26 minutes in all for $b\le16$ on six cores, and 5\,449\,s for the rotation with simplified adders at $b=17$
on ten.

The comparison with the textbook adder in Sec.~\ref{sec:adds} runs the same enumeration for the full-adder, simplified and fused cores
and counts basis inputs on which the gradient register does not end in $y+T(c)$ with phase $+1$. For every
$b=2,\dots,14$ the count is zero for the full-adder core and exactly 3/8 of all inputs (3/4 of those with $q'=1$) for the
simplified and fused cores, while condition (d) holds on every input of all three. For the fused core this follows from the
calculation in Appendix~\ref{app:givens-proof}: the gradient register changes only in its low $b-2$ bits, by
$s(M_{\mathrm{low}}-2^{b-2}c)$, which equals $T(c)$ modulo $2^b$ only when $-c\equiv m_{b-2}+2qv_{b-1}\pmod4$.

\section{Walk-step reconstruction}\label{app:walk}

Low \emph{et al.} give the Toffolis and qubits of each subroutine of a DFTHC+BLISS+SA walk step as formulas in $N$,
$(R,B,C)$, the bit precisions and the block sizes $2^{k}$ of the QROAM lookups (their App.~B\,7 and Tables~VII--IX); their
Table~III works them out for FeMoco-54. The Givens rotations enter as $4(N-1)\brot$ per SEL call, $8(N-1)\brot$ per step
(two SEL calls, each with $U_{\mathrm{Rot}}$ and $U_{\mathrm{Rot}}^\dagger$). The version of record revised two formulas
of the preprint: the inner QROAM output gains a sign qubit, $b_2=\lceil\log(B+1)\rceil+b_{k2}+1$ (their Eq.~(B36) and
Table~VIII, against $\lceil\log(B+1)\rceil+b_{k2}$ in the preprint), and the inner index register no longer stores $c$,
which removes $\lceil\log C\rceil$ qubits (their Eq.~(B39)); their Table~III still uses the preprint's values. Their
released code~\cite{Low2025code} implements the two versions as the classes \texttt{SpinFreeData\_V1} and
\texttt{SpinFreeData}, with $\lceil\log_2B\rceil$ in place of $\lceil\log_2(B+1)\rceil$, which differ only when $B$ is a
power of two, and without the sign qubit in the qubit count; the select cost is the same in both. We transcribe the
preprint's formulas and the two classes (Table~\ref{tab:walkfe2s2}) and, for each system, scan $\brot\in[10,20]$,
$b_{\mathrm{coeff}}\in[5,16]$ and four QROAM exponents in $[0,7]$, about $5\times10^{5}$ combinations, for exact agreement
with the published Toffolis and qubits. Seven of the eight entries of their Table~V rebuild exactly with
\texttt{SpinFreeData\_V1}; the eighth, CPD1, whose entry is unchanged from the preprint and which their driver script builds
without the V1 option, rebuilds with \texttt{SpinFreeData}. All six entries of their Table~XI, also unchanged from the
preprint, rebuild exactly with \texttt{SpinFreeData}; for FeMoco-76 there, with $B=64$, only its $\lceil\log_2B\rceil$
reproduces the entry. Each rebuild has the stated bit precisions and is the only one in the scan, except for \FeS{} in
their Table~XI, where two sets of block exponents give the same costs. The preprint's Table~5~\cite{Low2025} rebuilds with
the preprint's formulas, which also reproduce the totals of their Table~III, 9\,997 Toffolis and 1\,131 qubits for
FeMoco-54 with the parameters stated there. The fused rotation changes only the $8(N-1)\brot$ term, to
$4(N-1)(2\brot-5)$.

The logical qubit count is the sum of persistent registers, $2N$ system qubits and the largest temporary register. The
persistent registers are the uniform-state register, $\lceil\log_2L\rceil+2$; the alias-sampling output, $b_1+b_{k1}$;
the inner QROAM output and indices, $(\lceil\log_2B\rceil+b_{k2})+b_{k2}+\lceil\log_2R\rceil+2$ in
\texttt{SpinFreeData\_V1}; the inner uniform-state register, $\lceil\log_2(B+1)\rceil+2$; the rotation angles,
$(N-1)\brot$; and the phase-gradient register, $\brot$. For \FeS{} these are $9+29+32+6+285+15=376$. The largest
temporary is the outer alias QROAM's, $(\lceil\log_2L\rceil-2)+2^{k_1}b_1+\lceil\log_2\lceil L/2^{k_1}\rceil\rceil
=5+36+6=47$, so the total is $376+40+47=463$, as published. The full scan results and reconstructed parameters are
available from the authors on request, together with the verification and analysis scripts.

\begin{table*}[t]
\caption{\label{tab:walkfe2s2}One DFTHC+BLISS+SA walk step for \FeS{}, from the formulas of Low \emph{et al.}'s released
code (\texttt{SpinFreeData\_V1}) with $(N,R,B,C)=(20,14,15,5)$, $b_{k1}=b_{k2}=b_{\mathrm{coeff}}=11$, $\brot=15$ and
QROAM exponents $(k_1,k_2,k_4,k_5)=(1,3,1,3)$; $L=N+RC=90$, $b_1=\lceil\log_2L\rceil+b_{k1}=18$,
$b_2=\lceil\log_2B\rceil+b_{k2}+1=16$. Outer PREP and the reflections run once per step, inner PREP, RPREP and SEL
twice.}
\begin{ruledtabular}
\footnotesize\setlength{\tabcolsep}{3pt}
\begin{tabular}{llr|llr}
Subroutine & Toffolis per step & \FeS{} & Subroutine & Toffolis per step & \FeS{} \\
\hline
Outer uniform state & $8\lceil\log_2L\rceil$ & 56 & Inner QROAM inv., swaps & $2\cdot2^{k_4}(B+1)$ & 64 \\
Outer alias QROAM, unary & $\lceil L/2^{k_1}\rceil$ & 45 & Inner uniform states & $16\lceil\log_2(B+1)\rceil$ & 64 \\
Outer alias QROAM, swaps & $2^{k_1}b_1$ & 36 & RPREP angle QROM, unary & $2(N+RB)$ & 460 \\
Outer alias inv., compare & $b_1$ & 18 & RPREP erasure fix-up & $2(R+B)$ & 58 \\
Outer alias inv., unary & $\lceil L/2^{k_5}\rceil$ & 12 & SEL Givens rotations & $8(N-1)\brot$ & 2\,280 \\
Outer alias inv., swaps & $2^{k_5}$ & 8 & SEL spin controlled swaps & $4N$ & 80 \\
Inner QROAM, unary & $2RC\lceil(B+1)/2^{k_2}\rceil$ & 280 & SEL Majorana control & $2\cdot7$ & 14 \\
Inner QROAM, swaps & $2\cdot2^{k_2}b_2$ & 256 & Reflection $T_2$ & $\lceil\log_2(B+1)\rceil+b_{k2}+1$ & 16 \\
Inner QROAM inv., compare & $2b_2$ & 32 & Walk reflection & $\lceil\log_2L\rceil{+}\lceil\log_2(B{+}1)\rceil{+}b_{k1}{+}b_{k2}{+}2$ & 35 \\
Inner QROAM inv., unary & $2(N+\lceil RC/2^{k_4}\rceil)$ & 110 & & & \\
\hline
\multicolumn{2}{l}{Total (Low \emph{et al.}: 3\,924)} & 3\,924 & \multicolumn{2}{l}{With fused rotations, $4(N-1)(2\brot-5)=1\,900$} & 3\,544 \\
\end{tabular}
\end{ruledtabular}
\end{table*}

\section{Sparse \FeS{} Hamiltonian}\label{app:sparse}

With $a_j=(\gamma_{2j}+i\gamma_{2j+1})/2$, the spin-orbital Hamiltonian
$\sum h_{pq}a^\dagger_{p\sigma}a_{q\sigma}+\tfrac12\sum(pq|rs)a^\dagger_{p\sigma}a^\dagger_{r\tau}a_{s\tau}a_{q\sigma}$,
read from the FCIDUMP file with PySCF~\cite{pyscf}, is expanded into sorted Majorana monomials, with signs from reordering
and $\gamma^2=1$. Each monomial is one Pauli string up to a phase for any Jordan--Wigner ordering, so $L$ and $\lambda$
do not depend on the ordering. Two checks: the coefficients of two- and four-Majorana monomials are purely imaginary and
real (residual $5\times10^{-18}$), and a random 15$\alpha$/15$\beta$ determinant has the same energy from the expansion and
from the integrals ($2\times10^{-12}\Ha$). The count of terms above the cut is 41\,276 at $0.99\times10^{-4}\Ha$, 41\,052 at
$10^{-4}\Ha$ and 40\,736 at $1.01\times10^{-4}\Ha$, so the agreement with Gu \emph{et al.} pins the cut.

\section{Surface-code model}\label{app:cost}

We transcribe the model of Lee \emph{et al.}~\cite{Lee2021} as released in OpenFermion~\cite{OpenFermion}, which Low
\emph{et al.} state they use: surface-code patches~\cite{Fowler2012} at physical error rate $p=10^{-3}$ and a $1\,\mu$s
cycle, $\lceil1.5\,n_{\mathrm{log}}\rceil$ patches of $2(d+1)^2$ physical qubits for $n_{\mathrm{log}}$ logical qubits,
two-level AutoCCZ magic-state factories~\cite{GidneyFowler2019} labelled by their code distances $(\ell_1,\ell_2)$, a
per-patch per-cycle failure $0.1\,(100p)^{(d+1)/2}$, and a total failure budget of 0.1. With $n_{\mathrm F}=4$ factories each Toffoli
takes $\tau/n_{\mathrm F}$, where $\tau$ is the factory's time per CCZ state, and the layout minimizes physical qubits times run time
over factory types and odd distances. For fixed-hardware comparisons the factory type, factory count and $d$ chosen for
Low \emph{et al.}'s circuit are kept, and run time and failure are recomputed for ours.

The factories that appear are AutoCCZ(15,23), with a footprint of 138\,240 physical qubits, $\tau=115\,\mu$s and a failure
of $1.5\times10^{-10}$ per CCZ state; AutoCCZ(15,25), 151\,424, $125\,\mu$s and $6.3\times10^{-11}$; AutoCCZ(17,25),
237\,952, $125\,\mu$s and $1.1\times10^{-11}$; and AutoCCZ(17,27), 188\,160, $135\,\mu$s and $2.0\times10^{-12}$. A run
with $n_{\mathrm{Tof}}$ Toffolis takes $n_{\mathrm{Tof}}\tau/n_{\mathrm F}$, uses $\lceil1.5\,n_{\mathrm{log}}\rceil\times2(d+1)^2+n_{\mathrm F}\times$
(footprint) physical qubits, and fails with probability $n_{\mathrm{Tof}}p_{\mathrm{CCZ}}+0.1(100p)^{(d+1)/2}\times$
(patches) $\times$ (cycles). For FeMoco-76 the model selects four AutoCCZ(17,25) at $d=27$: 31.25\,$\mu$s per Toffoli,
8.71\,h, 4.37 million qubits and failure 0.079 for Low \emph{et al.}'s $1.00\times10^{9}$ Toffolis. Our qubit count is
2.9\% below their 4.5 million and our run time 1.3\% above their 8.6\,h, which their text gives together with the
preprint's $9.99\times10^{8}$ Toffolis; we did not find the source of these differences, and all comparisons are made
inside the model. Within the footprint of their FeMoco-76 layout the model's fastest layout is 7.52\,h with five
AutoCCZ(17,27) factories for their circuit and 5.21\,h with six AutoCCZ(15,25) for ours, but only because the shorter run
lets the smaller factory meet the failure budget, at 0.098; we use no result that depends on this threshold. The fastest layouts within a fixed footprint search factory counts 1--80,
all factory types and odd $d\le59$, with a run time of $\sum\max(t_r,\tau/n_{\mathrm F})$ per Toffoli and reaction time
$t_r=10\,\mu$s.

\bibliography{refs}

\end{document}